\documentclass[12pt]{article}
\def\be{\begin{equation}}
\def\ee{\end{equation}}
\def\bea{\begin{eqnarray}}
\def\eea{\end{eqnarray}}
\def\d{{\bf d}}
\def\p{{\bf p}}
\def\x{{\bf x}}
\def\y{{\bf y}}
\def\la{\lambda}

\def\t{\tilde}

\begin{document}

\begin{titlepage}
\bigskip
\rightline{}
\rightline{hep-th/0206228}
\bigskip\bigskip\bigskip\bigskip
   \centerline{\Large \bf {Instability of Spacelike and Null}}
   \bigskip
\centerline{\Large \bf {Orbifold Singularities}}
    \bigskip\bigskip
     \bigskip\bigskip

\centerline{\large Gary T. Horowitz${}^*$ and Joseph Polchinski${}^\dagger$}
\bigskip\bigskip
   \centerline{\em ${}^*$Department of Physics, UCSB, Santa Barbara, CA 93106}
   \centerline{\em ${}^\dagger$Kavli Institute for Theoretical Physics,
   UCSB, Santa Barbara, CA 93106}
    \bigskip\bigskip
\begin{abstract}
Time dependent orbifolds with
spacelike or null singularities have recently been studied as
simple models of cosmological singularities. We show that their apparent
simplicity is an illusion: the introduction of a single particle causes the
spacetime to collapse to a strong curvature singularity (a Big Crunch),
even in regions arbitrarily far from the particle.
\end{abstract}
   \end{titlepage}

\baselineskip=18pt

\section{Introduction}

Understanding the physics at cosmological
singularities has long been a
challenge for string theory. One of the main open questions is whether
time (as we know it) simply begins and ends, or do
quantum effects produce a kind of bounce, with a well defined
semi-classical spacetime on the other side. (One of the earliest
attempts to apply string theory to cosmology, the pre-Big Bang
scenario \cite{Veneziano:2000pz}, requires a bounce.) In order to study this
question, it is natural to start with the simplest examples of solutions
with cosmological singularities: time dependent orbifolds. These are
quotients of flat spacetime by boosts (or combinations of boosts and
rotations). Since the spacetimes are locally flat, they are exact
classical solutions to string theory and string propagation
is relatively easy to study. Given that
the singularities of ordinary orbifolds are harmless
in string theory, one might expect the same would be true here.
However, arguments were given many years ago
that this would not be the case \cite{hsteif}.

Recently, interest has been
refocussed on this question
by the ekpyrotic cosmology~\cite{ekpyrotic}, which
requires~\cite{singular}
that the universe contract to a singularity and then
reexpand.
The ekpyrotic singularity is spacelike and corresponds to
orbifolding by a  simple boost. The cyclic universe model
\cite{Steinhardt:2001vw} has a similar orbifold singularity.
It has been argued that
from
some points of view this singularity is rather
gentle~\cite{KOSST,totu}, while from other points of view it is not
\cite{Nekrasov:2002kf}\footnote{It was noted in~\cite{Neil} that metric
perturbations become large near the orbifold singularity.}.
One difficulty with this orbifold is that
the resulting spacetime has closed timelike curves.
   As a warmup, Liu,
Moore, and
Seiberg (LMS)~\cite{LMS} have studied the orbifold by
a null
boost~\cite{hsteif,simon}.  This has no closed
timelike curves, though it
does have a submanifold with closed null curves,  and a null singularity.
It also has the advantage of being
supersymmetric.  LMS study four-string amplitudes in
this
space, and finds that they are nonsingular except for
exceptional momenta. They do not study the backreaction and
so are not able to conclude whether there is a bounce or not.

In this note we will reexamine this question, and will
argue that in fact
these orbifold singularities are highly unstable. Given their
application to both M theory and string theory, we will consider
general $D$ dimensional orbifolds. For a simple boost, the identification
acts on a two dimensional subspace, so there are $D-2$ flat transverse
directions. For the null orbifold, the identification involves three
dimensions, so there are $D-3$ transverse directions.
We will show that the addition
of even a {\it
single} particle, localized in the transverse directions,
causes the entire universe to
collapse into a spacelike
singularity.
The basic mechanism is the amplification of
the energy of particles in the collapsing universe, a
point that has been
emphasized by many others going back to
ref.~\cite{hsteif}.  We will
analyze the effects of this first in the context of
general relativity,
finding evidence for the collapse, and then argue that
this effect can be
seen in the string amplitudes of ref.~\cite{LMS}.
Unfortunately, this does not resolve the question of
what happens at cosmological singularities in string theory.
It just shows that orbifolds are not really any simpler than
other cosmological models. One must learn to deal with the
strong curvature effects.

While this was being
written, several related papers appeared.
Ref.~\cite{Lawrence:2002aj} shows that a {\it homogeneous}
energy distribution in the transverse directions causes the null
orbifold singularity to become spacelike.
Refs.~\cite{LMS2,fabing} extend the analysis of LMS to the
nonsingular orbifold in which the null boost is accompanied by a
spacelike shift, and include remarks about the backreaction that
overlap with ours.  We should also note other discussions of Lorentzian
orbifolds, of flat and curved spacetimes~\cite{misc}.

\section{General relativity argument for the Big Crunch}

In this section we use general relativity to study the effect of
introducing a single particle into the orbifolds with timelike and null
singularities. The basic idea is to go to the covering space,
and view the single particle on the
orbifold as an infinite collection of particles in Minkowski spacetime
all related by the appropriate boost. If the gravitational interaction
of these particles produce a singularity, the same must be true
for a single particle on the orbifold. Due to the large boosts involved,
the rest mass of the particle can be neglected. We will therefore consider
massless particles.

\subsection{Preliminaries}

We begin with the basic system of two massless
particles with momenta $P^\mu$ and $\t P^\mu$ in $D$ spacetime
dimensions. Let $b$ be the impact parameter in the center of
mass frame. Since the center of mass energy squared is
$s = - 2 P\cdot \t P$,
the condition that
the particles approach within their Schwarzschild radius
and form a black
hole is
\be
G \sqrt{s} > b^{D-3}  \label{schwarz}
\ee
up to coefficients of order one.  This heuristic
condition can be
justified in general relativity as follows.
The gravitational field of a single
massless particle is given by a shock wave
called the
Aichelberg-Sexl metric \cite{Aichelburg:1970dh}. For $D>4$, the metric is
\be\label{as}
ds^2 = -2 dy^+ dy^- + d\y^2 + {\mu\over (\y\cdot \y)^{(D-4)/2}} 
\delta(y^+) (dy^+)^2
\ee
where $\mu$ is proportional to the energy of the particle. In $D=4$, the
metric is  similar, with the power law dependence in the last term
replaced by a log. The curvature is concentrated on a null plane, and
falls off as one moves away from the particle. The high energy collision
of two particles can be described in the center of mass frame
by superposing two such shock waves
in the past.
This is possible, despite the nonlinearities of general relativity, because the
spacetime is flat in front of each shock wave. After the particles collide,
the exact solution is not known. But if the impact parameter is small enough,
there are trapped surfaces in the
spacetime when the particles collide \cite{penrose}.
It follows from the singularity theorems that a singularity
must form.
Assuming cosmic censorship, the singularity must be inside a black
hole. In four spacetime dimensions, a trapped surface has been found
\cite{Eardley:2002re} provided
the impact parameter satisfies the bound (\ref{schwarz}). In higher dimensions,
a trapped surface has been found for $b=0$ \cite{Eardley:2002re}.
The size of the trapped surface is
of order the Schwarzschild radius for the given energy, so one again expects a
black hole to form whenever $b$ satisfies (\ref{schwarz}), but this
has not yet been rigorously shown.

It will be useful to have a general formula for the impact parameter $b$ in
the center of mass frame, given two null geodesics
\be
X^\mu(\la) = P^\mu \la + a^\mu\ , \quad\t X^\mu(\t \la) =
\t P^\mu \t \la + \t a^\mu\ .
\ee
This is easily obtained as follows.  Projecting
each trajectory into the subspace orthogonal to $P^\mu + \t P^\mu $, which is at
constant time in the center of
mass frame, we get
\be
X^\mu(\la)\rightarrow {1\over 2} (P -\t P)^\mu \la + a^\mu -
(P + \t P)^\mu { a\cdot (P + \t P) \over 2P\cdot \t P}\ ,
\ee
\be
\t X^\mu(\t \la)\rightarrow - {1\over 2} (P -\t P)^\mu \t \la + \t a^\mu
-(P + \t P)^\mu\,{ \t a\cdot (P + \t P)
\over 2P\cdot \t P}\ .
\ee
Now take the difference between these two trajectories and compute the
norm. The norm depends on $\la +\t \la$, and minimizing with respect to this
parameter yields
\be\label{impact}
b^2 = Y^2 - \frac{2 (P \cdot Y)\, (\t P \cdot Y)}{P
\cdot \t P}
\ee
where $Y^\mu= a^\mu - \t a^\mu$. Since $a^\mu$ and $\t a^\mu$
are arbitrary points along the geodesics,
one can set $Y^\mu$ equal to the difference between {\it any} pair
of points on the respective trajectories,
$Y^\mu = X^\mu(\la) - \t X^\mu(\t \la)$.
This does not change the impact parameter because
(\ref{impact}) is invariant under shifting $Y^\mu$ by any multiple of
$P^\mu$ or $\t P^\mu$.

\subsection{Two examples}

We now consider our first example:  two dimensional
Minkowski spacetime quotiented by a boost, times a flat transverse space.
The spacetime is
\be
ds^2 = -2dy^+ dy^- + d\y^2
\ee
with the identification
\be\label{boostorb}
(y^+, y^-, \y) \equiv ( e^{n\alpha} y^+,  e^{-n\alpha} y^-, \y)
\ee
for all integers $n$ and some constant $\alpha$.
This orbifold has  fixed points at $y^+ = 0,\ y^-=0$ which is a spacelike
singularity.  It consists of four regions, two with closed timelike
curves.\footnote
{To avoid this, one might consider just the past and future
wedges.  Since this is not a true orbifold
(it is a
quasiorbifold in the terminology of ref.~\cite{panic})
there are no methods as yet for studying such a space in string theory,
or even for showing that it exists as a solution.}
As mentioned above, we will study the effect of adding a
single massless particle to this
orbifold by going to the
covering space where one has the original particle together
with an infinite number of boosted images. The momentum of the $n^{\rm th}$
image is related to the original momentum $(p^+, p^-, \p)$ by
$( e^{n\alpha} p^+, e^{-n\alpha} p^-,\p)$.
   On the
covering space, a sufficient condition for the
formation of a black hole,
and therefore a spacelike singularity, is that any pair of images
satisfy the condition~(\ref{schwarz}).
Consider say the original particle and its $n^{\rm th}$ image. Their
center of mass energy
grows like $s\sim \cosh n\alpha$. To compute the impact parameter, we use
(\ref{impact}). It is
convenient to choose the affine parameter $\la$ to vanish at $X^+=0$,
so $a^+=0$. Then $Y^2$ only depends on the transverse components of $a$
which are independent of $n$. The $n$ dependent contribution to
$P\cdot Y$ is $- P^+ a^- (1-e^{-n\alpha})$ for the original particle and
$-P^+ a^- (e^{n\alpha} -1)$ for the $n^{\rm th}$ image.
This implies that for large $|n|$, the impact parameter
is independent of $n$.

It follows that for large
$n$ the condition~(\ref{schwarz}) for formation of a
black hole is always
satisfied.  Moreover, by taking large enough $n$, the
Schwarzschild radius
\be
R_{\rm s}^{D-3} \sim  G  (p^+ p^-)^{1/2} e^{n\alpha/2}
\label{rs}
\ee
becomes arbitrarily large, so that the black hole
occupies all of space. In other words, the entire spacetime
ends in a curvature singularity and one has a
Big Crunch. It has often been noted that the spacetime inside a black hole
is analogous to an (anisotropic) cosmology with a big crunch. When
the black hole is arbitrarily large so there is no spacetime outside the
horizon, this analogy becomes exact.

We now show that the same thing happens for the null orbifold, which is
described by
\be
ds^2 = -2 dx^+ dx^- + dx^2 + d\x^2
\ee
with the identification
\be\label{nullorb}
   X \equiv (x^+, x, x^-, {\bf x})\
\stackrel{\sim}{=}\ X_n \equiv (x^+, x
+ nx^+, x^- + nx + n^2x^+/2, {\bf x})
\ee
Note that  the $D-3$ transverse coordinates $\x$
are not affected by the identification. The fixed points are at
$x^+=0, x=0$ and form a null singularity.
There is a similar identification of the momenta
\be
   P \equiv (p^+, p, p^-, {\bf p})\
\stackrel{\sim}{=}\ P_n \equiv (p^+, p
+ np^+, p^- + np + n^2p^+/2, {\bf p})\ .
\ee
%\bea
%&& X \equiv (x^+, x, x^-, {\bf x})\
%\stackrel{\sim}{=}\ X_n \equiv (x^+, x
%+ nx^+, x^- + nx + n^2x^+/2, {\bf x})\ ;\nonumber\\
%&& P \equiv (p^+, p, p^-, {\bf p})\
%\stackrel{\sim}{=}\ P_n \equiv (p^+, p
%+ np^+, p^- + np + n^2p^+/2, {\bf p})\ .
%\eea
(Note $X_0 \equiv X$ and $P_0 \equiv P$ by definition.)

As before, a massless particle in this orbifold can be described
by an infinite collection of particles on the covering space. If we
focus on one particle with $X(\la) = P\la +a$ and its $n^{\rm th}$ image,
the center of mass energy is
\be
s_n = -2 P \cdot P_n = 2 p^+ p^-_n + O(n) =
n^2 (p^+)^2 \ ,
\ee
and so the center of mass energy grows as $n$. To compute the impact
parameter, we again choose the affine parameter to vanish at $X^+=0$, so
$a^+=0$.
   One quickly
verifies that $Y^2$ is independent of $n$, and that in
the second term
of~(\ref{impact}) both numerator and denominator are of
order $n^2$, so that the
impact parameter does not grow with $n$.
Once again, one produces black holes with arbitrarily large size by taking
$n$ sufficiently large.

Thus, in both cases,  a single
particle induces a gravitational collapse everywhere in spacetime.
Essentially, the gravitational shock wave
of the particle wraps
infinitely many times through the identified space and
intersects itself
infinitely many times, and successive intersections
are increasingly
energetic.

One might wonder whether the intersections at smaller
$n$ might introduce
nonlinear effects that would prevent the formation of
larger black holes at
larger $n$.  This should not be the case, because the
trapped surface
corresponding to larger $n$ is at larger radius,
spacelike separated from
the region of nonlinearity.  In fact, the combined
effect of multiple
shocks goes the other way: if we consider the total
center of mass energy
of the first $n$ images of each particle, then one
finds that (in, e.g., the null orbifold)
\be
   s_{\rm tot} \propto n^4\ ,\quad b \propto n^0 \
,\quad
R_{\rm s}^{D-3} \sim  G n^2 p^+
\label{big}
\ee
so $R_{\rm s}$ grows even more quickly with $n$.

The collision of two shock waves changes the spacetime only to the future
of their intersection. We now  ask whether the orbifold singularity
indeed lies to the future of the intersection of two shocks.
A null particle $P^\mu \la + a^\mu$ generates a
shock wave consisting of all  points $x^\mu$ satisfying $P\cdot
(x - a)=0$. The $n^{th}$ boosted image produces the shock wave 
$P_n\cdot (x - a_n)=0$. Since $P\cdot a = P_n \cdot a_n$ by Lorentz invariance,
the intersection of the two  shocks must satisfy $(P_n - P)\cdot x =0$.
For the null orbifold, this implies
\begin{equation}
(np+n^2 p^+/2) x^+ = n p^+ x\ .
\end{equation}
Depending on the sign of $x$, one can take $n$ either  
positive or negative to insure that $x^+ <0$. Note that for large $|n|$,
$x^+ = O(1/|n|)$. Since every point with $x^+ >0$ lies to the future of
every point with $x^+ <0$ \cite{LMS},
it is clear that the entire region to the future
of the orbifold singularity lies to the future of the intersection of
the shocks.

For the boost orbifold (\ref{boostorb}) the condition $(P_n - P)\cdot y =0$
yields
\begin{equation}\label{intersect}
p^+ y^- = p^- y^+ e^{-n \alpha} \ ,
\end{equation}
Since $y^+$ and $y^-$ always have the same sign, the intersection never
occurs in the region of closed timelike curves. If we substitute 
(\ref{intersect})
into the condition for a single shock, $P\cdot(y-a)=0$, we obtain
\begin{equation}
(e^{-n \alpha}  + 1)p^- y^+ =  {\bf p} \cdot {\bf y} -P\cdot a
\end{equation}
If the right hand side is negative, the orbifold singularity
lies to the future of the intersection of the shocks for all images $n$.
If the right hand side is positive, it lies to the past. However even in this
case, it is
not possible for a particle to propagate into the future cone
before it encounters strong curvature.  If we ask when the trajectory
$P^\mu \la + a^\mu$ collides with the image shock $P_n\cdot (y-a_n)=0$,
a short calculation shows that the affine parameter $\lambda$ is
negative (for positive $n$) and so the collision occurs in the
past cone.

It is somewhat counterintuitive that a single particle
can change the geometry
at arbitrary distance.  In particular, it might appear to violate causality.
To see that this is not the case, and make the above result more plausible,
we present another argument which leads to the same conclusion.
For definiteness, we will consider the null orbifold.
Suppose we start with a massive particle rather than a massless one (the
above argument still goes though essentially unchanged.)
This particle produces a gravitational perturbation which at large distances
is described by the linearized
Schwarzschild solution:
\be\label{schw}
ds^2 = -dt^2 + dr^2 + r^2 d\Omega_{D-2}^2 + \left({r_0\over r}\right)^{D-3}
\left[dt^2 + {1\over D-3}(dr^2 + r^2 d\Omega_{D-2}^2)\right]
\ee
On the covering space, we have an infinite number of particles
which produce a field at large transverse distance which is just the sum of
the Schwarzschild perturbation of each one. In terms of the above
null coordinates,
$t= (x^+ + x^-)/\sqrt 2, \ r^2 = (x^+ - x^-)^2/2 + x^2 + {\bf x}^2$. It is
easy to show that
this sum diverges for $x^+=  x=0$ and any transverse
separation $\x^2$. Since we start with a particle at $r=0$, it clearly
passes through the fixed point surface $x^+=0, x=0$, so all of its
images agree at this point.
Thus the radial distance $r$ from each is the same
and one has an infinite number of copies of a perturbation of strength
$(r_0/r)^{D-3}$. Actually the divergence is stronger than this, since the
particles have a  relative boost. If one takes the perturbation
(\ref{schw}) in the rest frame of the $n^{\rm th}$ image, and translates back
into the original coordinates, one finds e.g. $(dx^-)^2 \rightarrow (dx^-
+ ndx+n^2 dx^+/2)^2$ so different components of the perturbation pick up
extra powers of $n$. Of course, once the perturbation becomes large, one
can no longer trust the linearized approximation, but this clearly shows
why strong gravitational fields can arise from a single particle even at
large transverse distance\footnote{For $D=5$, the sum of the image perturbations
diverges at all times, even long before the orbifold
singularity. However, this is just
a gauge artifact~\cite{thankNeil}.}.

One can also show that the perturbation becomes large
even when the initial particle has $x\ne 0$ at $x^+=0$.
From our earlier argument we know that the minimum
separation between the original particle and its $n^{\rm th}$ image is
independent of $n$.  This minimum separation may be reached at different times
for  different images. Nevertheless, focusing on just the initial particle
and one image, at the time of minimum separation, the field
at a large transverse distance will include the sum of two Schwarzschild
perturbations with essentially the same $r$. Since certain components
of the perturbation are multiplied by powers of $n$, for any transverse
distance, one can choose $n$ large enough so that the perturbation
is larger than one.

In the above, we considered a generic particle with $p^+ \ne 0$. The situation
is different if one introduces a particle with $p^+=0$, so the trajectory
lies in a null surface of constant $x^+$. Since the total momentum must be
null, the only nonzero component can be $p^-$. In our first orbifold,
obtained by quotienting by a simple boost, the description of this
particle  in the
covering space is an infinite series of parallel particles converging to
$x^+=0$ with increasing energy. The gravitational shock waves of these
particles do not intersect, and the entire spacetime
can be viewed as a time dependent pp wave. However, since the energy of the
particles diverge as $x^+ \rightarrow
0$ the pp wave becomes singular there, much like the singular plane waves
studied in \cite{Horowitz:sr}.
For the null orbifold (\ref{nullorb}), its
clear that the image trajectories in the covering space are simply translated
in $x$ direction and have exactly the same momentum. Parallel null particles
never form a black hole. The gravitational backreaction is described by
a single shock wave at one value of $x^+$, with a profile that is
a linear superposition of the shock wave for each particle.  In this one
nongeneric case there is no singularity.

\subsection{Generalizations}

There is a generalization of the null orbifold which regulates the
singularity. If one adds a commuting spacelike shift to the null
boost, then there are no
fixed points. The identification is now
\be
(x^+, x, x^-, {\bf x})\ \stackrel{\sim}{=}\ (x^+, x
+ nx^+, x^- + nx + n^2x^+/2, {\bf x} +
n\d)\ .
\ee
This has been called a ``null brane" in the literature
\cite{Figueroa-O'Farrill:2001nx,simon}.
Geometrically,
there is a compact direction which starts at infinite radius in the
past, contracts down
to a size given by $\d$ and then expands out to infinity. What happens
if one adds a single massless particle to a null brane? The extra shift
does not affect the center of mass energy of two image particles.
However it has
the important effect that the impact parameter now grows linearly with $n$.
The result depends\footnote{We thank H. Liu and N. Seiberg for a discussion
on this point.} on the total dimension $D$. Consider first a
chain of identical point masses,  each with
mass $M$.
If the separation $|\d|$
is smaller than $R_s$ where $R_s^{D-3}\sim GM$, there will be a
cylindrical event horizon surrounding the masses, i.e., a black
string. The transverse size of the black string
is $R_{tr}^{D-4} \sim GM/|\d|$. This transverse size can also be
obtained as follows.
The total mass clearly grows
linearly with distance
along the chain. But in $D>4$ dimensions, the Schwarzschild radius grows
more slowly. So at most a finite number of masses
contribute to form a
black hole. Setting $R_s^{D-3} = GMn$ equal to $(n\d)^{D-3}$,
solving for $n$ and substituting back in, one finds that $R_s$ agrees
with the transverse size of the black string.

The only difference between this chain and the
null brane is that, in the center of mass frame, the energy of each particle
in the chain grows linearly with $n$, so the total mass grows like $n^2$.
It follows that for $D=5$, no black hole forms for large shift, and an
infinite mass black hole forms for small shift, as before.
In $D>5$, the situation is different. A large shift again produces no
singularities, but even a small shift will produce only a finite size
black hole. Outside this black hole, there will not be a big crunch.
The spacetime will approach the null brane at large distances. The
size of the black hole can be estimated as follows. A black hole
of size $R_s$ will contain $n$ images where $R_s = n|\d|$. Since the
mass of the $n$ images is of order $n^2 p^+$, we have
\be
R_s ^{D-3} = G n^2 p^+ = (n\d)^{D-3} \ .
\ee
Solving for $n$ yields
\be
R_s = (G p^+/ \d^{2})^{1/(D-5)} \ .
\ee
One sees clearly that as the shift $\d$ goes to zero, the size of the
black hole grows to infinity.

One can do the same thing for the orbifold with a spacelike singularity
(\ref{boostorb}).
By adding a commuting shift to the standard boost, one avoids the singularity
\cite{Cornalba:2002fi}.
In this case, 
since the center of mass energy grows exponentially and the separation
only grows linearly, a single particle will still produce a Big
Crunch 
for any finite shift\footnote{We thank Liu and Seiberg, and Cornalba and Costa
for
pointing out an incorrect statement in an earlier version of this paper.}.

The orbifold (\ref{boostorb}) can be written in the form
\be
ds^2 = -dt^2 + t^2 d\phi^2 + d\y^2
\ee
where $\phi$ is periodic. If we identify $\phi$ with $-\phi$, we obtain a
model of two ``end of the world" branes colliding. This is the geometry of
the cyclic universe model \cite{Steinhardt:2001vw}. This clearly has the same
instability as we discussed above. However,
in the cyclic universe, one might expect quantum fluctuations to stop the
branes from hitting at exactly the same time everywhere \cite{Stein}.
Can this also
regulate the singularity and avoid the
instability? This is very unlikely for two reasons. First, our
argument is  local. We considered just a neighborhood of a single particle
and showed that its interaction with its images produce black holes of
unbounded size. We did not have to assume that the circle was shrinking
down to zero size everywhere at exactly the same time. Second,
if the quantum fluctuations lead to classical perturbations,
then they will classically grow and produce curvature singularities even
without introducing extra particles.
For example, one can easily verify that any metric of the form
\be
ds^2 = -dt^2 + [t+ f(\y)]^2 d\phi^2 + d\y^2
\ee
has a curvature singularity when $t+f(\y) =0$, unless $f$ is a linear
function.

\section{String theory argument for the Big Crunch}

So far, our analysis has been strictly in the context of general
relativity.  One might hope that the situation would be better in string
theory --- that stringy physics in these Lorentzian orbifolds would be
nonsingular, just as it is in Euclidean orbifolds.  However, this is unlikely.
The singularity involves the formation of an arbitrarily large black hole,
with Schwarzschild radius much larger than the string scale.  At these
distances string theory should go over to general relativity.  LMS
\cite{LMS} have
calculated string scattering amplitudes in the null orbifold, and so we can
look for the expected breakdown of perturbation theory in these.

Let us first ask how one would detect the onset of black hole formation in the
tree level $2\to 2$ string amplitude.  First in general
relativity, the covariant scattering amplitude is of order
\be
{\cal A} \sim G s^2/t
\label{gramp}
\ee
with $s$ the center-of-mass energy squared and $t$
the momentum transfer squared.  Fourier transforming with
respect to the $D-2$
transverse dimensions, and including a factor of
$s^{-1}$ to convert from
covariant to canonical normalization of states, yields
a dimensionless amplitude
\be
{\delta} \sim \frac{Gs}{b^{D-4}}  \label{amp}
\ee
where $b$ is the impact parameter.  In string theory, this is modified at $b^2
\stackrel{<}{\sim} \alpha' \ln s$ by the logarithmic spreading of the
string~\cite{suss}, but this is much smaller than the Schwarzschild radius.
There is also an amplitude for the strings to become excited, but this is
again small at large radius.  Thus the general relativistic result~(\ref{amp})
extends to string theory.

The dimensionless amplitude~(\ref{amp}) becomes of order one at $b \propto
s^{1/(D-4)}$, which at high energy is much larger than the Schwarzschild radius
$b \propto s^{1/(2D-6)}$.  Thus perturbation theory breaks down long before
black holes form.  There is a simple reason for this.  At macroscopic
distances and energies, a classical description of the gravitational field is
valid.  One can think of this as the exchange of many gravitons, which is a
high order ladder graph, so indeed perturbation theory in this sense has broken
down. Since there is a classical description, there should be a way to sum the
large terms in perturbation theory.  This is the eikonal
approximation~\cite{ACV,eikonal}.  Essentially, the large amplitude
exponentiates to give the S-matrix
\be
S = \exp(2i \delta + \ldots)\ .
\ee
The phase is large, but the nontrivial physical effect comes only through the
dependence of the phase on $b$.  A measure of the magnitude of this is the
scattering angle
\be
\theta \sim s^{-1/2} \frac{d\delta}{db} \sim \frac{Gs^{1/2}}{b^{D-3}}\ .
\label{angle}
\ee
We see that $\theta \sim 1$ is the criterion for black hole formation.  This
agrees with the classical analysis of scattering
of ultrarelativistic particles: the energy at which a black hole forms is of
the same order as that where the scattering angle becomes large.\footnote
{The reader might be concerned that the signature for black hole formation
is a scattering angle of order one, which is highly suppressed at high energy
at string tree level, but what the
angle~(\ref{angle}) actually represents is the effect of many soft
scatterings.}

Now let us look for this effect in the string amplitudes in the
null orbifold geometry.  To compare with the $2\to 2$ tree amplitude
in LMS, we consider a slightly different situation from before --- the
interaction of one particle with the images of another, rather than with its
own images (to see the latter effect, we would need to look at string loop
amplitudes).  In general relativity the argument in the previous section 
still goes through;  for
the $n^{\rm th}$ image, the center of mass energy grows as $n$, while the
minimum separation does not, and so a black hole of arbitrarily large radius
forms.

For simplicity let us analyze the
kinematics in the case that $P$ and $\t P$ are purely in
the $+$ direction; one can check that the analysis extends directly to
more generic momenta, and to massive external particles.
Then
\be
P_0 = (p^+, 0, 0, {\bf 0})\ ,\quad \t P_n = (\t p^+,
n\t p^+, n^2 \t p^+/2,
{\bf 0}) \ .
\ee
If these exchange a momentum
\be
K =  (k^+, 0, k^- , {\bf k})\ ,
\ee
then the mass shell conditions $(P_0 + K)^2 = (\t P_n - K)^2 = 0$
imply that for large $n$
\be
k^- = \frac{{\bf k}^2}{2p^+}\ ,\quad k^+ = -\frac{2
{\bf k}^2}{n^2 \t p^+}\ .
\ee
The key point is that $k^+$ is very small, of order $1/n^2$.  This is just
the region where LMS noted that their amplitude diverges.  Thus we interpret
this divergence as an indication of the breakdown of perturbation theory
due to the onset of black hole formation.  The kinematics above corresponds,
in the notation of LMS ($p_1 \equiv p$, $p_2 \equiv \tilde p$) to
$L_s = n^2 p^+_1 p^+_2 \Rightarrow q_+^2 = n^2 p^+_1 p^+_2/(p_1^+ + p_2^+)$;
$L_t - 4/\alpha' \sim {\bf k}^2$; $p_3^+ - p_1^+ = k^+$.  In this regime the
amplitude $A$ in LMS 6.16 reduces to the general relativistic form
(\ref{gramp}), and for $\Delta J = 0$ the phase factor in LMS 6.16 is
negligible.  Thus the general relativistic analysis of this regime is not
altered.

Our general relativistic analysis was in the spirit of the inheritance
principle: for untwisted states, tree level amplitudes descend from
amplitudes on the covering space.  It is not obvious that this is
valid here.  Multiple graviton exchange is a multiloop process, even though the
eikonal approximation allows it to be summed up in terms of classical
general relativity.  What our analysis has ignored is the exchange of
winding states (which also would not be seen in the tree level string amplitude
considered above).  These states become light near $x^+ = 0$ where the black
hole is forming, and so it is conceivable that they
qualitatively change the process.  Note that their effect is limited
by causality, because they are heavy until just before the instant $x^+ = 0$.

This is hardly the last word on this subject, but we can summarize our
conclusions as follows.  The best reason for believing that a bounce occurs
in this context is the resemblance of these spacetimes to Euclidean
orbifolds.  However, an application of orbifold technology shows that in
fact these singularities are unstable toward the formation of
singularities of a more terminal sort.  The orbifold singularities are no
better (or worse) than the spacelike curvature singularities of black holes,
and so we must still understand the physics of these in string theory.

\vskip 1in
\centerline{\bf Acknowledgements}
\vskip 1cm
It is a pleasure to thank Hong Liu, Nati Seiberg and Neil Turok for stimulating
discussions. The work of GH was supported in part by NSF grant
PHY-0070895. The work of JP was supported in part by NSF grants PHY99-07949
and PHY00-98395.


\begin{thebibliography}{10}
\baselineskip=14pt

%\cite{Veneziano:2000pz}
\bibitem{Veneziano:2000pz}
G.~Veneziano,
``String cosmology: the pre-big bang scenario,''
arXiv:hep-th/0002094.
%%CITATION = HEP-TH 0002094;%%

\bibitem{hsteif}
G.~T.~Horowitz and A.~R.~Steif,
``Singular string solutions with nonsingular initial
data,''
Phys.\ Lett.\ B {\bf 258}, 91 (1991).
%%CITATION = PHLTA,B258,91;%%

\bibitem{ekpyrotic}
J.~Khoury, B.~A.~Ovrut, P.~J.~Steinhardt and N.~Turok,
``The ekpyrotic universe: colliding branes and the
origin of the hot big
bang,'' Phys.\ Rev.\ D {\bf 64}, 123522 (2001)
[arXiv:hep-th/0103239].
%%CITATION = HEP-TH 0103239;%%

\bibitem{singular}
R.~Kallosh, L.~Kofman, A.~D.~Linde and A.~A.~Tseytlin,
``BPS branes in cosmology,''
Phys.\ Rev.\ D {\bf 64}, 123524 (2001)
[arXiv:hep-th/0106241].
%%CITATION = HEP-TH 0106241;%%

%\cite{Steinhardt:2001vw}
\bibitem{Steinhardt:2001vw}
P.~J.~Steinhardt and N.~Turok,
``A cyclic model of the universe,''
arXiv:hep-th/0111030.
%%CITATION = HEP-TH 0111030;%%

\bibitem{KOSST}
J.~Khoury, B.~A.~Ovrut, N.~Seiberg, P.~J.~Steinhardt
and N.~Turok,
``From big crunch to big bang,''
Phys.\ Rev.\ D {\bf 65}, 086007 (2002)
[arXiv:hep-th/0108187].
%%CITATION = HEP-TH 0108187;%%

\bibitem{totu}
A.~J.~Tolley and N.~Turok,
``Quantum fields in a big crunch / big bang
spacetime,''
arXiv:hep-th/0204091.
%%CITATION = HEP-TH 0204091;%%

%\cite{Nekrasov:2002kf}
\bibitem{Nekrasov:2002kf}
N.~A.~Nekrasov,
``Milne universe, tachyons, and quantum group,''
arXiv:hep-th/0203112.
%%CITATION = HEP-TH 0203112;%%

\bibitem{Neil}
J.~Khoury, B.~A.~Ovrut, P.~J.~Steinhardt and N.~Turok,
``Density perturbations in the ekpyrotic scenario,''
arXiv:hep-th/0109050.
%%CITATION = HEP-TH 0109050;%%

\bibitem{LMS}
H.~Liu, G.~Moore and N.~Seiberg,
``Strings in a time-dependent orbifold,''
arXiv:hep-th/0204168.
%%CITATION = HEP-TH 0204168;%%



\bibitem{simon}
J.~Simon,
``The geometry of null rotation identifications,''
arXiv:hep-th/0203201.
%%CITATION = HEP-TH 0203201;%%

%\cite{Lawrence:2002aj}
\bibitem{Lawrence:2002aj}
A.~Lawrence,
``On the instability of 3D null singularities,''
arXiv:hep-th/0205288.
%%CITATION = HEP-TH 0205288;%%

\bibitem{LMS2}
H.~Liu, G.~Moore and N.~Seiberg,
``Strings in time-dependent orbifolds,''
arXiv:hep-th/0206182.
%%CITATION = HEP-TH 0206182;%%

\bibitem{fabing}
M.~Fabinger and J.~McGreevy,
``On smooth time-dependent orbifolds and null singularities,''
arXiv:hep-th/0206196.
%%CITATION = HEP-TH 0206196;%%

\bibitem{misc}
F.~Larsen and F.~Wilczek,
``Resolution of cosmological singularities,''
Phys.\ Rev.\ D {\bf 55}, 4591 (1997)
[arXiv:hep-th/9610252];\\
%%CITATION = HEP-TH 9610252;%%
G.~T.~Horowitz and D.~Marolf,
``A new approach to string cosmology,''
JHEP {\bf 9807}, 014 (1998)
[arXiv:hep-th/9805207];\\
%%CITATION = HEP-TH 9805207;%%
V.~Balasubramanian, S.~F.~Hassan, E.~Keski-Vakkuri and A.~Naqvi,
``A space-time orbifold: A toy model for a cosmological singularity,''
arXiv:hep-th/0202187;\\
%%CITATION = HEP-TH 0202187;%%
S.~Elitzur, A.~Giveon, D.~Kutasov and E.~Rabinovici,
``From big bang to big crunch and beyond,''
arXiv:hep-th/0204189;\\
%%CITATION = HEP-TH 0204189;%%
B.~Craps, D.~Kutasov and G.~Rajesh,
``String propagation in the presence of cosmological singularities,''
arXiv:hep-th/0205101;\\
%%CITATION = HEP-TH 0205101;%%
E.~J.~Martinec and W.~McElgin,
``Exciting AdS orbifolds,''
arXiv:hep-th/0206175.
%%CITATION = HEP-TH 0206175;%%

\bibitem{Aichelburg:1970dh}
P.~C.~Aichelburg and R.~U.~Sexl,
``On the gravitational field of a massless particle,''
Gen.\ Rel.\ Grav.\  {\bf 2}, 303 (1971).
%%CITATION = GRGVA,2,303;%%

\bibitem{penrose}
R. Penrose, unpublished (1974);
P.~D.~D'Eath and P.~N.~Payne,
``Gravitational radiation in high speed black hole collisions. 1.
Perturbation treatment of the axisymmetric speed of light collision,''
Phys.\ Rev.\ D {\bf 46}, 658 (1992).
%%CITATION = PHRVA,D46,658;%%

%\cite{Eardley:2002re}
\bibitem{Eardley:2002re}
D.~M.~Eardley and S.~B.~Giddings,
``Classical black hole production in high-energy collisions,''
arXiv:gr-qc/0201034.
%%CITATION = GR-QC 0201034;%%

\bibitem{panic}
A.~Adams, J.~Polchinski and E.~Silverstein,
``Don't panic! Closed string tachyons in ALE
space-times,''
JHEP {\bf 0110}, 029 (2001)
[arXiv:hep-th/0108075].
%%CITATION = HEP-TH 0108075;%%

\bibitem{thankNeil}
N. Turok, private communication.

%\cite{Horowitz:sr}
\bibitem{Horowitz:sr}
G.~T.~Horowitz and A.~R.~Steif,
``Strings in strong gravitational fields,''
Phys.\ Rev.\ D {\bf 42}, 1950 (1990).
%%CITATION = PHRVA,D42,1950;%%


%\cite{Figueroa-O'Farrill:2001nx}
\bibitem{Figueroa-O'Farrill:2001nx}
J.~Figueroa-O'Farrill and J.~Simon,
``Generalized supersymmetric fluxbranes,''
JHEP {\bf 0112}, 011 (2001)
[arXiv:hep-th/0110170].
%%CITATION = HEP-TH 0110170;%%



%\cite{Cornalba:2002fi}
\bibitem{Cornalba:2002fi}
L.~Cornalba and M.~S.~Costa,
``A new cosmological scenario in string theory,''
arXiv:hep-th/0203031.
%%CITATION = HEP-TH 0203031;%%

\bibitem{Stein}
P. Steinhardt, private communication.

\bibitem{suss}
L.~Susskind,
``String theory and the principles of black hole complementarity,''
Phys.\ Rev.\ Lett.\  {\bf 71}, 2367 (1993)
[arXiv:hep-th/9307168].
%%CITATION = HEP-TH 9307168;%%

\bibitem{ACV}
D.~Amati, M.~Ciafaloni and G.~Veneziano,
``Superstring collisions at Planckian energies,''
Phys.\ Lett.\ B {\bf 197}, 81 (1987);
%%CITATION = PHLTA,B197,81;%%
``Classical and quantum gravity effects from Planckian energy superstring
collisions,'' Int.\ J.\ Mod.\ Phys.\ A {\bf 3}, 1615 (1988);
%%CITATION = IMPAE,A3,1615;%%
``Effective action and all order gravitational eikonal at Planckian energies,''
Nucl.\ Phys.\ B {\bf 403}, 707 (1993).
%%CITATION = NUPHA,B403,707;%%

\bibitem{eikonal}
G.~'t Hooft,
``Graviton dominance in ultrahigh-energy scattering,''
Phys.\ Lett.\ B {\bf 198}, 61 (1987);\\
%%CITATION = PHLTA,B198,61;%%
I.~J.~Muzinich and M.~Soldate,
``High-energy unitarity of gravitation and strings,''
Phys.\ Rev.\ D {\bf 37}, 359 (1988);\\
%%CITATION = PHRVA,D37,359;%%
H.~Verlinde and E.~Verlinde,
``Scattering at Planckian energies,''
Nucl.\ Phys.\ B {\bf 371}, 246 (1992)
[arXiv:hep-th/9110017];\\
%%CITATION = HEP-TH 9110017;%%
D.~Kabat and M.~Ortiz,
``Eikonal quantum gravity and Planckian scattering,''
Nucl.\ Phys.\ B {\bf 388}, 570 (1992)
[arXiv:hep-th/9203082].
%%CITATION = HEP-TH 9203082;%%

\end{thebibliography}
\end{document}